\documentclass[12pt]{article}
\usepackage{amsfonts}
\usepackage[square,sort&compress, numbers, comma]{natbib}
\usepackage{amsmath,amssymb,array,graphicx}
\usepackage{graphicx}
\usepackage[dvips]{color}
\usepackage{subfigure}
\usepackage{multirow}
\usepackage{natbib}

\newcommand{\beq}{\begin{equation}}
\newcommand{\eeq}{\end{equation}}
\newcommand{\bey}{\begin{eqnarray}}
\newcommand{\eey}{\end{eqnarray}}
\newcommand{\nn}{\nonumber}

\def\ds{\displaystyle}

\def\p{{\partial}}

\usepackage{setspace}
\usepackage{sectsty}
\sectionfont{\large}
\begin{document}

\newpage
\title{Pricing Parisian down-and-in options}
\author{Song-Ping Zhu, Nhat-Tan Le, Wen-Ting Chen and Xiaoping Lu
\\School of Mathematics and Applied Statistics, \\University of Wollongong,\\
NSW 2522, Australia. }

\date{}
\maketitle
%\vspace{-1.5cm}
\begin{abstract}
In this paper, we price American-style Parisian down-and-in call options under the Black-Scholes framework. Usually, pricing an American-style option is much more difficult than pricing its European-style counterpart because of the appearance of the optimal exercise boundary in the former. Fortunately, the optimal exercise boundary associated with an American-style Parisian knock-in option only appears \emph{implicitly} in its pricing partial differential equation (PDE) systems, instead of \emph{explicitly} as in the case of an American-style Parisian knock-out option. We also recognize that the ``moving window'' technique developed for pricing European-style Parisian up-and-out options can be adopted to price American-style Parisian knock-in options as well. In particular, we obtain a simple analytical solution for American-style Parisian down-and-in call options and our new formula is written in terms of four double integrals, which can be easily computed numerically.

\end{abstract}

\smallskip\noindent

\vspace{-0.8cm}
\smallskip\noindent
\vspace{-0.6cm}


\vspace{-0.2cm}
\section{Introduction}
\vspace{-0.3cm}

Barrier options are common path-dependent options traded in financial markets. One possible reason is that this kind of options provides a more flexible and cheaper way for hedging and speculating than vanilla options because the option buyers only pay a premium for scenarios they perceive as likely. The ``one touch" breaching barrier, however, may have an undesirable feature of suddenly losing all proceeds (knock-out options) or suddenly receiving the embedded option (knock-in options) if the price of the underlying momentarily touches the asset barrier, no matter how briefly the breaching occurs. This opens up the possibility of market practitioners deliberately manipulating the underlying asset to force the cancelation or activation of the option. To partially remedy such a drawback, Parisian options are introduced, with a unique feature that the underlying asset has to continually stay above or below the asset barrier for a prescribed amount of time before the knock-out or knock-in feature is activated. This extended trigger clause can also be found in some derivative contracts, such as callable convertible bonds and executive warrants \cite{dai2004knock}. It is also worthwhile to note that Parisian options can be a useful tool in corporate finance \cite{bernard2005new}.

%Similar to barrier options, Parisian options can be classified into two types in terms of actions being triggered: Parisian knock-out options and Parisian knock-in options. The holder of the former starts with a vanilla option, but will lose this vanilla option if the knock-out feature is activated. In contrast, the holder of the latter has nothing to start with, but will immediately receive a vanilla option if the knock-in feature is triggered. Moreover, to clarify the exercise style of the embedded vanilla option, which is ``knocked in" or ``knocked out", an extra term ``American-style" or ``European-style" is used in the name of the Parisian options. For instance, the term ``American-style" in the name of an ``American-style Parisian knock-in options" refers to the American-style exercise of the vanilla option obtained after the knock-in feature is activated.

Like the relationship between a vanilla American option and its European counterpart, the valuation problem of American-style Parisian options, in general, is much more difficult than that of their European-style counterparts. While a closed-form analytical solution of the latter has already been found by \cite{zhu2013pricing}, a closed-form solution of the former only exists for the perpetual knock-in case (cf. \cite{chesney2006american}).
In this paper, an explicit analytical solution for American-style Parisian knock-in call options is found after adopting the ``window technique" developed in \cite{zhu2013pricing}. Our solution procedure can be extended to find analytical solutions of other types of Parisian knock-in options.
 %It is anticipated that our solution will contribute to the literature in both theoretical and practical aspects. Theoretically, our solution procedure can be extended to find closed-form analytical solutions of other types of Parisian knock-in options. Practically, with a growing demand for trading exotic options in today's finance industry, our solution procedures may lead to the development of pricing formulae for other exotic derivatives, such as the Edokko options introduced by \citet{fujita2002edokko}, which are generalizations of both Parisian and delayed barrier options.

The paper is organized as follows. In Section 2, we introduce the PDE systems governing the price of Parisian down-and-in call options. In Section 3 the solution procedure is presented. Conclusion is given in the last section.

\section{The PDE systems}
\vspace{-0.3cm}
%The extra difficulty of pricing American-style Parisian options, in comparison with their European counterparts, mainly stems from the complexity of the determination of the optimal exercise price. For example, in the case of an American-style Parisian knock-out call option,
%the option holder, on the one hand, has the incentive to wait for the asset price to further increase, hoping for more financial gain when exercising the option. On the other hand, he/she also has to bear more risks of losing the option altogether if the asset continue to stay above or below the asset barrier for ``too long'' and eventually the ``knock out" mechanism is triggered. As a result,  the optimal exercise boundary is now a three-dimensional (3-D) surface, instead of a two-dimensional (2-D) curve for the case of an American vanilla option, and this is the primary source of difficulty for pricing American-style Parisian knock-out options.
One does not need to deal directly with the optimal exercise boundary in the evaluation of American-style Parisian knock-in options. This is because holders of an American-style Parisian knock-in option cannot do or decide anything until the option is activated.  Moreover, once the ``knock-in'' feature is activated, the optimal exercise boundary of the option is the same as that of the embedded vanilla American option, the calculation of which has been thoroughly studied in the literature \cite{ kim1990analytic,zhu2006exact,zhu2006new,zhu2007calculating,cox1979option,muthuraman2008moving,garcia2003convergence,jacka1991optimal,gao2000valuation, longstaff2001valuing}. In other words, the optimal exercise boundary does not appear explicitly before the option is knocked-in, and is already determined after the option being knocked-in.
This suggests that the valuation of American-style Parisian knock-in options should be similar to that of European-style Parisian knock-in options and a simple analytical solution can be achieved.

%In this section, the PDE systems governing the price of an American Parisian down-and-in call option are established under the Black-Scholes framework. As we expected and will see later, the free boundary does not appear explicitly in the PDE systems, although the option is of American style.

%In this section, based on the same financial arguments in \cite{zhu2013pricing}, we can elegantly simplify the pricing domain of an American-style Parisian Down-and-in Call option. We then establish the PDE systems, under the Black-Scholes framework, governing the option price in the reduced domain.

Theoretically speaking, an American Parisian down-and-in call option will be knocked in and become the embedded American vanilla call if the underlying asset continually stays below the barrier $\bar{S}$ for a prescribed time period $\bar{J}$. Otherwise, the Parisian down-and-in call option will be expired worthless.
%Moreover, before the ``knock-in" feature is triggered, any early exercise of the option yields a zero value.

For some extreme values of the ``barrier", one can easily observe an American Parisian down-and-in call option becomes worthless or degenerates to either a one-touch barrier option or a vanilla option. For example, if $\bar{J}$ approaches zero, the option will be immediately ``knocked in'' once the underlying touches the barrier from the above, which is the same as the specification of a one-touch barrier call option with down-and-in feature. Similarly, it can be deduced that if $\bar{J}$ is greater than the option life, $T$, or $\bar{S}$ approaches zero, the option values nothing. On the other hand, if $\bar{S}$ approaches infinity and $\bar{J}$ is less than T, the option price should be the same as that of the associated American call as the knock-in feature will be surely activated.

For other non-degenerate cases, the price of an American Parisian down-and-in call option is, however, not trivial. It depends on the underlying price $S$, the current time $t$ and the barrier time $J$, in addition to other parameters such as the volatility rate $\sigma$, risk-free interest rate $r$ and the expiry time $T$. We now assume that the underlying asset $S$ with a continuous dividend yield $D$ follows a lognormal Brownian motion given by
\beq dS=(r-D)Sdt+\sigma S dZ,\label{1.2.1}\eeq
where $Z$ is a standard Brownian motion.

Based on the same financial arguments in \cite{zhu2013pricing}, the pricing domains of those non-degenerated cases can be elegantly reduced as
 \bey &&I:~ \{\bar{S}\leq S < \infty ,~0\leq t\leq T-\bar{J},~ J=0\},\nn\\
&&II:~ \{0\leq S\leq\bar{S},~J\leq t\leq J+T-\bar{J},~0\leq J\leq \bar{J}\}.\nn\eey
Let $V_1(S,t)$ and $V_2(S,t,J)$ denote the option prices in the region $I$ and $II$, respectively. Based on the definition of the option, the continuity condition of the option price and the ``option Delta", it can be shown that the option price should satisfy {{ (cf.\cite{zhu2013pricing, Haber})}}
\begin{equation*}
\mathcal{A}_1\begin{cases}
\ds \frac{\p V_1}{\p t}+ \mathbb{L}V_1=0, \\
\vspace{.15cm} V_1(S,T-\bar{J})  =0, \\
\vspace{.15cm} \ds \lim_{S\rightarrow \infty}V_1(S,t) = 0,\\
\vspace{.15cm} \ds V_1(\bar{S},t)=V_2(\bar{S},t,0),
\end{cases}
\quad
\mathcal{A}_2
\begin{cases}
\ds \frac{\p V_2}{\p t}+ \frac{\p V_2}{\p J}+\mathbb{L}V_2=0, \\
\vspace{.15cm} V_2(S,t,\bar{J}) =C_{A}(S, t),\\
\vspace{.15cm} \ds V_2(0,t,J) = 0,\\
\vspace{.15cm} \ds V_2(\bar{S},t,J)=V_2(\bar{S},t,0), 0 \leq J <\bar{J}
\end{cases}
\end{equation*}
%\vspace{-0.8cm}
\beq
\text{connectivity condition} : \ds \frac{\p V_1}{\p S}(\bar{S},t)=\frac{\p V_2}{\p S}(\bar{S},t,0),
\label{1.3.2}\eeq
where $\mathcal{A}_1$ is defined on $t\in [0,T-\bar{J}],~S\in[\bar{S}, \infty)$, and $\mathcal{A}_2$ is defined on $t\in [J,T-\bar{J}+J],~J\in[0,\bar{J}],~S\in[0, \bar{S}]$, operator $\ds\mathbb{L}=\frac{\sigma^2 S^2}{2}\frac{\p^2}{\p S^2}+(r-D)S\frac{\p}{\p S}-rI$, with $I$ being the identity operator.
%
%
%\begin{equation*}
%\hspace{-3.5cm}\mathcal{A}_1\begin{cases}
%\ds {\frac{\p V_1}{\p t}+ \mathbb{L}V_1=0, 0<t < T-\bar{J}, \bar{S} < S < + \infty}\\
%\vspace{.15cm} {V_1(S,T-\bar{J})  =0},\\
%\vspace{.15cm} {\ds \lim_{S\rightarrow +\infty}V_1(S,t) = 0}, \\
%\vspace{.15cm} {\ds \lim_{S\rightarrow \bar{S}}V_1(S,t)=V_2(S,t,0)},
%\end{cases}
%\end{equation*}
%\begin{equation*}
%\mathcal{A}_2
%\begin{cases}
%\ds{ \frac{\p V_2}{\p t}+ \frac{\p V_2}{\p J}+\mathbb{L}V_2=0,  0<t < T-\bar{J}, - \infty< S <\bar{S}, 0 < J <\bar{J} }\\
%\vspace{.15cm} { V_2(S,t,\bar{J}) =C_{A}(S, t)}, \\
%\vspace{.15cm} \ds {V_2(0,t,J) = 0}, \\
%\vspace{.15cm} \ds { \lim_{S\rightarrow \bar{S}} V_2(S,t,J)=V_2(S,t,0),}
%\end{cases}
%\end{equation*}
%
%%\vspace{-0.8cm}
%\beq
%\text{connectivity condition} : {\ds \frac{\p V_1}{\p S}(\bar{S},t)=\frac{\p V_2}{\p S}(\bar{S},t,0), \forall 0 \leq t \leq T-\bar{J}},
%\label{1.3.2}\eeq
%{where operator $\ds\mathbb{L}=\frac{\sigma^2 S^2}{2}\frac{\p^2}{\p S^2}+(r-D)S\frac{\p}{\p S}-rI$, with $I$ being the identity operator.}

   One can observe that the above PDE systems are quite similar to those governing a European-style Parisian up-and-out call option. However, there are still several key differences. Firstly, it is obvious that the pricing domain of a Parisian down-and-in option is reversed from that of its up-and-out counterpart. Secondly, the knock-in feature makes the ``terminal condition", with respect to $J$, become non-homogeneous in $\mathcal{A}_2$. This is because the option price is equal to that of the associated American call option, denoted by $C_{A}(S, t)$, at the time $t$ it is ``knocked in". Finally, we have the homogeneous boundary condition in $\mathcal{A}_1$ when S becomes very large because in this case the option is never ``knocked in".

Albeit different, the above coupled PDE systems can be solved by adopting the ``moving window'' technique developed in \cite{zhu2013pricing}. In the next section, we shall briefly discuss the solution procedure.

\section{Solution of the coupled PDE systems}
{{We first replace the sum of the two partial derivatives appearing in $\mathcal{A}_2$, i.e., $\ds\frac{\p V_2}{\p t}+\frac{\p V_2}{\p J}$, by the directional derivative $\ds\sqrt{2}\frac{\p V_2}{\p l}$. As a result, the governing equation in the new coordinate system can be written as \beq \ds (\sqrt{2}\frac{\p }{\p l}+\mathbb{L})V_2(S,l;t)=0,\label{1.3.4}\eeq
where $t$ serves as a parameter, identifying the slide passing through the point $(\bar{S},t,0)$. Furthermore, the constant $\sqrt{2}$ can be absorbed by rescaling $l$, i.e., $\ds l^{'}=\frac{l}{\sqrt{2}}$, and (\ref{1.3.4}) becomes \beq \ds (\frac{\p }{\p l^{'}}+\mathbb{L})V_2(S,l^{'};t)=0,\eeq which is nothing but the BS equation!}}
\begin{figure}[h!]\center
  \includegraphics[width=.65\textwidth]{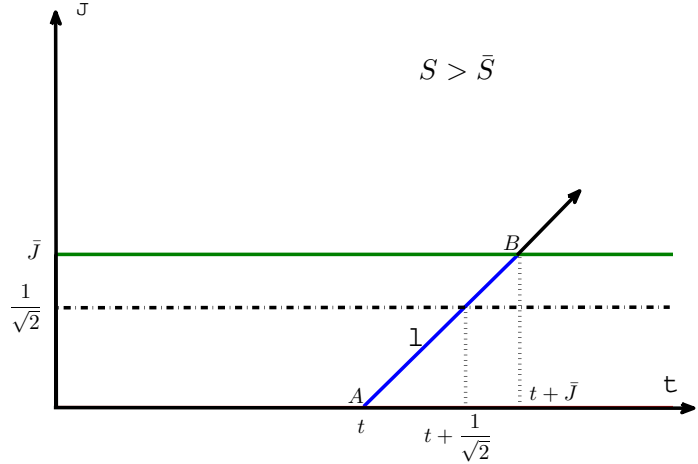}\\
  %\vspace{-1cm}
  \caption{Projection of a typical slide on the $t-J$ plane}
\end{figure}

{{A 2-D diagram demonstrating the above coordinate transformation is shown in Fig 1, where the line $AB$ denotes the projection on the $t-J$ plane of the slide passing through the point $(\bar{S},t,0)$. From this figure, one can also observe that the solution $V_2(S,l';t)$ in the new coordinate system is equal to $\ds V_2(S,t+l',l')$ in the original pricing domain $II$. Particularly, let $l' = \bar{J}$, we have $V_2(S,\bar{J};t)= V_2(S,t+\bar{J},\bar{J} ) = C_{A}(S, t+\bar{J})$.}}

%\vspace{-0.5cm}Following the method of \cite{zhu2013pricing} with the same notations, the 3-D PDE systems (\ref{1.3.2}) can be further simplified to the following two 2-D PDE systems
Once all the boundary conditions in the new coordinate system are worked out accordingly, the 2-D PDE systems governing the price of an American-style Parisian down-and-in call option can be summarized as
\begin{equation*}
\mathcal{A}_1\begin{cases}
\ds \frac{\p V_1}{\p t}+ \mathbb{L}V_1=0, \\
\vspace{.15cm} V_1(S,T-\bar{J})  =0, \\
\vspace{.15cm} \ds \lim_{S\rightarrow \infty}V_1(S,t) = 0, \\
\vspace{.15cm} \ds  V_1(\bar{S},t)=W(t),
\end{cases}
\quad
\mathcal{A}_2\begin{cases}
\ds \frac{\p V_2}{\p l^{'}}+\mathbb{L}V_2=0, \\
\vspace{.15cm} V_2(S,\bar{J};t) =C_{A}(S, t + \bar{J}), \\
\vspace{.15cm} \ds V_2(0,l^{'};t) = 0, \\
\vspace{.15cm} \ds V_2(\bar{S},l^{'};t)=W(t+l^{'}),
\end{cases}
\end{equation*}

\beq \text{connectivity condition} : \ds \frac{\p V_1}{\p S}(\bar{S},t)=\frac{\p V_2}{\p S}(\bar{S},0;t),
\label{1.3.5}\eeq
where $W(t)$ denotes the time-dependent function as $V_2(\bar{S},0;t)$. This unknown function $W(t)$ that provides the coupling between the two PDE systems also needs to be solved as part of the solution. Here, $\mathcal{A}_1$ is defined on $t\in [0,T-\bar{J}],~S\in[\bar{S}, \infty)$, and $\mathcal{A}_2$ is defined on $t\in [0,T-\bar{J}],~l'\in[0,\bar{J}],~S\in[0, \bar{S}]$.

To solve the newly established pricing system (\ref{1.3.5}) effectively, we shall first non-dimensionalize all variables. In addition to the dimensionless variables introduced in \cite{zhu2013pricing}, we further introduce: $$\bar{J}^{'}=\frac{\sigma^2\bar{J}}{2},~T^{'}=\frac{\sigma^2 T}{2},~C_{A}(S,t) = K C^{'}_{A}(x,\tau +\bar{J'}), \tau=(T-\bar{J}-t)\frac{\sigma^2}{2}$$

With all primes and tildes dropped from now on, the dimensionless PDE system reads:
\begin{equation}
\mathcal{A}_1\begin{cases} \label{I}
\ds \frac{\p V_1}{\p \tau}=\mathbb{L}V_1, \\
\vspace{.15cm} \ds V_1(x,0)  = 0, \\
\vspace{.15cm} \ds \lim_{x\rightarrow \infty}V_1(x,\tau) = 0, \\
\vspace{.15cm} \ds V_1(\bar{x},\tau)=W(\tau),
\end{cases}
\quad
\mathcal{A}_2\begin{cases}
\ds \frac{\p V_2}{\p l}=\mathbb{L}V_2, \\
\vspace{.15cm} V_2(x,0;\tau) = C_{A}(x,\tau), \\
\vspace{.15cm} \ds \lim_{x\rightarrow -\infty}V_2(x,l;\tau) = 0, \\
\vspace{.15cm} \ds V_2(\bar{x},l;\tau)=W(\tau-\bar{J}+l),
\end{cases}
\end{equation}
\vspace{-0.4cm}
\beq \text{connectivity condition} : \ds\frac{\p V_1}{\p x}(\bar{x},\tau)=\frac{\p V_2}{\p x}(\bar{x},\bar{J};\tau),
\label{1.3.6}\eeq
where $\mathcal{A}_1$ is defined on $\tau\in [0,T-\bar{J}],~x\in[\bar{x}, \infty)$, and $\mathcal{A}_2$ is defined on $\tau\in [0, T- \bar{J}],~l\in[0,\bar{J}],~x\in(-\infty, \bar{x}]$, operator $\ds\mathbb{L}=\frac{\p^2}{\p x^2}+k\frac{\p }{\p x}-\gamma I$ with $k$ being equal to $\gamma-q-1$.

By applying the Laplace transform technique as well as the Green function method, we obtain
\beq \ds V_1(x,\tau)=\int^{\tau}_0 W(s)g_1(x,\tau-s)ds,\label{1.3.7}\eeq
\vspace{-0.5cm}where $$ g_1(x,\tau)=\frac{x-\bar{x}}{2\sqrt{\pi}\tau^{\frac{3}{2}}}e^{-(\frac{k^2}{4}+\gamma)
\tau-\frac{(x-\bar{x})^2}{4\tau}-\frac{k}{2}(x-\bar{x})}.$$
%f(z)&=& V_{BS}(z,\bar{J}).\nn\eey
Similarly, $V_2$ can be solved as
\beq \ds V_2(x,l;\tau)=F(x,l;\tau)+\int^{l}_0 W(\tau-\bar{J}+s)g_2(x,l-s)ds, \label{1.3.8}\eeq
where $g_2(x,l) = -g_1(x,l)$ and $$ F(x,l;\tau)=\int^{\bar{x}}_{-\infty}\frac{1}{2\sqrt{\pi l}}e^{-\frac{k}{2}(x-z)-(\frac{k^2}{4}+\gamma)l}
[e^{-\frac{(x-z)^2}{4l}}-e^{-\frac{(x+z-2\bar{x})^2}{4l}}]C_{A}(z,\tau)dz.$$

Now, applying the connectivity condition (\ref{1.3.6})  to (\ref{1.3.7}) and (\ref{1.3.8}), we obtain the integral equation governing $W(\tau)$ as
\beq\int^{\tau}_0 W(s)\frac{\p g_1}{\p x}(x,\tau-s)ds|_{x=\bar{x}}=\ds\frac{\p F}{\p x}(x,\bar{J};\tau)|_{x=\bar{x}}+ \int^{\bar{J}}_{0} W(\tau-\bar{J}+s)\frac{\p g_2}{\p x}(x,\bar{J}-s)ds|_{x=\bar{x}},\eeq
which can be written as below after a new variable transform $\xi=\tau-\bar{J}+s$ is introduced
\beq\int^{\tau}_0 W(s)\frac{\p g_1}{\p x}(x,\tau-s)ds|_{x=\bar{x}}=\ds\frac{\p F}{\p x}(x,\bar{J};\tau)|_{x=\bar{x}}+ \int^{\tau}_{\tau - \bar{J}} W(\xi)\frac{\p g_2}{\p x}(x,\tau - \xi)d\xi|_{x=\bar{x}}.\label{1.3.10}\eeq
%\beq \ds\frac{\p F}{\p x}(x,\tau)|_{x=\bar{x}}+\int^{\tau}_0 W(s)\frac{\p g_1}{\p x}(x,\tau-s)ds|_{x=\bar{x}}=\int^{\tau}_{\tau - \bar{J}} W(\xi)\frac{\p g_2}{\p x}(x,\tau -\xi)ds|_{x=\bar{x}}.\label{1.3.10}\eeq
It can be observed that the left hand side of (\ref{1.3.10}) contains the information of $W(s)$ from the expiry ($\tau=0$) to the current time to expiry, $\tau$, while its right hand side involves the value of $W(\xi)$, with $\xi$ varying within $[\tau-\bar{J},\tau]$, which coincides with the projection of the ``slide'' (a plane is of $45^{\circ}$ angle to both of the plane $t=0$, and $J=0$) passing through $(\bar{S},\tau,0)$ on the plane $J=0$. As in \cite{zhu2013pricing}, we also name such a projection as a ``window". On the initial window, with $\tau$ varying within $[-\bar{J},0]$, the option price is the same as that of the standard American call and the value of $W(\tau)$ is defined by $W_0(\tau)=C_{A}(\bar{x},\tau+\bar{J})$.

By solving the integral equation (\ref{1.3.10}) with $\tau$ varying within $[0,\bar{J}]$, we obtain the value of W in the first window, denoted by $W_1(\tau)$, as
\bey W_1(\tau)&=&\int^{\bar{x}}_{-\infty}\frac{\bar{x}-z}{4\pi\bar{J}^{3/2}} e^{-\frac{(\bar{x}-z)^2}{4\bar{J}}-(\frac{k^2}{4}+\gamma)\bar{J}-\frac{k}{2}(\bar{x}-z)}
\int^{\tau}_0\frac{C_{A}(z,s)}{\sqrt{\tau-s}}e^{-(\frac{k^2}{4}+\gamma)(\tau-s)}dsdz\nn\\
&&+\frac{W_0(0)}{2}e^{-(\frac{k^2}{4}+\gamma)\tau}-\frac{e^{-(\frac{k^2}{4}+\gamma)\bar{J}}}{2\pi \sqrt{\bar{J}}}\int^{\tau}_0\frac{e^{-(\frac{k^2}{4}+\gamma)(\tau-s)}}{\sqrt{\tau-s}}W_0(s-\bar{J})ds\label{1.3.13}\\
&&-\frac{1}{\pi}\int^{\tau}_0 \frac{e^{-(\frac{k^2}{4}+\gamma)(\tau-s)}}{\sqrt{\tau-s}} \int^{\sqrt{\bar{J}}}_{\sqrt{s}}e^{-(\frac{k^2}{4}+\gamma)t^2}\big[(\frac{k^2}{4}+\gamma)W_0(s-t^2)+W^{'}_0(s-t^2)\big]dtds.\nn
\eey

It is quite interesting to observe that the last three terms in the above formula are identical to those in the corresponding formula for the European-style Parisian up-and-out call \cite{zhu2013pricing}. The differences are only in the first term, especially with the price of the standard American call appearing in the integrand.

Similar to the case in \cite{zhu2013pricing}, for a state point $(S,\tau,J)$, one can evaluate $W$ forwards, window by window, until the value at the required time $\tau$ is found. However, the determination of $W_{n+1}$, assuming that the option price on the $n$th window is known, is slightly different from that of $W_1$. In fact, in the new coordinate system $\tilde{\tau}=\tau-n\bar{J}$, solving $W_{n+1}(\tau)$ with the known option price on the $n$th window is equivalent to determining $U(\tilde{\tau})$ from the following PDE system:
\begin{equation}
\mathcal{B}_1\begin{cases} \label{II}
\ds \frac{\p V_1}{\p \tilde{\tau}}=\mathbb{L}V_1, \\
\vspace{.15cm} \ds V_1(x,0)  = f_n(x), \\
\vspace{.15cm} \ds \lim_{x\rightarrow \infty}V_1(x,\tilde{\tau}) = 0, \\
\vspace{.15cm} \ds V_1(\bar{x},\tilde{\tau})=U(\tilde{\tau}),
\end{cases}
\quad
\mathcal{B}_2\begin{cases}
\ds \frac{\p V_2}{\p l}=\mathbb{L}V_2, \\
\vspace{.15cm} V_2(x,0;\tilde{\tau}) = C_{A}(x,\tilde{\tau}), \\
\vspace{.15cm} \ds \lim_{x\rightarrow -\infty}V_2(x,l;\tilde{\tau}) = 0, \\
\vspace{.15cm} \ds V_2(\bar{x},l;\tilde{\tau})=U(\tilde{\tau}-\bar{J}+l),
\end{cases}
\end{equation}
\vspace{-0.8cm}
\beq \text{connectivity condition} : \ds \frac{\p V_1}{\p x}(\bar{x},\tilde{\tau})=\frac{\p V_2}{\p x}(\bar{x},\bar{J};\tilde{\tau}).
\eeq
Here $\ds f_n(x)=V_1(x,n\bar{J})=\sum^{n}_{i=1}\int^{i \bar{J}}_{(i-1)\bar{J}} W_i(s)g_1(x,n\bar{J}-s)ds$, $\ds\mathbb{L}=\frac{\p^2}{\p x^2}+k\frac{\p }{\p x}-\gamma I$, with $k$ being equal to $\gamma-q-1$. Moreover, $\mathcal{B}_1$ is defined on $\tilde{\tau}\in [0, \bar{J}],~x\in[\bar{x}, \infty)$, and $\mathcal{B}_2$ is defined on $\tilde{\tau}\in [0, \bar{J}],~l\in[0,\bar{J}],~x\in(-\infty, \bar{x}]$.

The non-homogeneous initial condition of the system $\mathcal{B}_1$ makes its solution procedure more complicated than that of $\mathcal{A}_1$. However, using the Laplace transform technique and the Green function method, we have managed to derive its solution as \beq \ds V_1(x,\tilde{\tau})=G(x,\tilde{\tau})+\int^{\tilde{\tau}}_0 { {U}}(s)g_1(x,\tilde{\tau}-s)ds,\eeq
\vspace{-0.5cm}where $$G(x,\tilde{\tau})=\int_{\bar{x}}^{+\infty}\frac{1}{2\sqrt{\pi \tilde{\tau}}}e^{-\frac{k}{2}(x-z)-(\frac{k^2}{4}+\gamma)\tilde{\tau}}
[e^{-\frac{(x-z)^2}{4\tilde{\tau}}}-e^{-\frac{(x+z-2\bar{x})^2}{4\tilde{\tau}}}]f_n(z)dz.$$
The corresponding integral equation governing $U(\tilde{\tau})$ is
\beq\ds\frac{\p G}{\p x}(x,\tilde{\tau})|_{x=\bar{x}}+\int^{\tilde{\tau}}_0 U(s)\frac{\p g_1}{\p x}(x,\tilde{\tau}-s)ds|_{x=\bar{x}}=\ds\frac{\p F}{\p x}(x,\bar{J};\tilde{\tau})|_{x=\bar{x}}+ \int^{\bar{J}}_{0} U(\tilde{\tau}-\bar{J}+s)\frac{\p g_2}{\p x}(x,\bar{J}-s)ds|_{x=\bar{x}},\label{1.3.11}\eeq
which can be solved as
\begin{multline*}
U(\tilde{\tau}) = \int_{\bar{x}}^{+\infty}\frac{e^{-\frac{k}{2}(\bar{x}-z)-(\frac{k^2}{4}+\gamma)\tilde{\tau}}}{2\sqrt{\pi\tilde{\tau}}}e^{-\frac{(\bar{x}-z)^2}{4\tilde{\tau}}}f_n(z)dz-\frac{e^{-(\frac{k^2}{4}+\gamma)\bar{J}}}{2\pi \sqrt{\bar{J}}}\int^{\tilde{\tau}}_0\frac{e^{-(\frac{k^2}{4}+\gamma)(\tilde{\tau}-s)}}{\sqrt{\tilde{\tau}-s}}{ {U}}_0(s-\bar{J})ds\\
\vspace{.5cm}+\frac{{ {U}}_0(0)}{2}e^{-(\frac{k^2}{4}+\gamma)\tilde{\tau}}+ \int^{\bar{x}}_{-\infty}\frac{\bar{x}-z}{4\pi\bar{J}^{3/2}}e^{-\frac{(\bar{x}-z)^2}{4\bar{J}}-(\frac{k^2}{4}+\gamma)\bar{J}-\frac{k}{2}(\bar{x}-z)} \int^{\tilde{\tau}}_0\frac{C_{A}(z,s)}{\sqrt{\tilde{\tau}-s}}e^{-(\frac{k^2}{4}+\gamma)(\tilde{\tau}-s)}dsdz\\
-\frac{1}{\pi}\int^{\tilde{\tau}}_0 \frac{e^{-(\frac{k^2}{4}+\gamma)(\tilde{\tau}-s)}}{\sqrt{\tilde{\tau}-s}} \int^{\sqrt{\bar{J}}}_{\sqrt{s}}e^{-(\frac{k^2}{4}+\gamma)t^2}\big[(\frac{k^2}{4}+\gamma){ {U}}_0(s-t^2)+{ {U}}^{'}_0(s-t^2)\big]dtds,
\end{multline*}
{{where $ U_0(\tilde{\tau}) = W_n(\tilde{\tau}+n\bar{J}), \forall \tilde{\tau} \in [-\bar{J},0]$}}.\\
Consequently, the analytical formula for $W_{n+1}(\tau)$ is
\begin{multline*}
W_{n+1}(\tau)=\int_{\bar{x}}^{+\infty}\frac{e^{-\frac{k}{2}(\bar{x}-z)-(\frac{k^2}{4}+\gamma)(\tau-n\bar{J})}}{2\sqrt{\pi(\tau-n\bar{J})}}e^{-\frac{(\bar{x}-z)^2}{4(\tau-n\bar{J})}}f_n(z)dz\\ +\frac{W_n(n\bar{J})}{2}e^{-(\frac{k^2}{4}+\gamma)(\tau-n\bar{J})}-\frac{e^{-(\frac{k^2}{4}+\gamma)\bar{J}}}{2\pi \sqrt{\bar{J}}}\int^{\tau}_{n\bar{J}}\frac{e^{-(\frac{k^2}{4}+\gamma)(\tau-s)}}{\sqrt{\tau-s}}W_n(s-\bar{J})ds\\
+\int^{\bar{x}}_{-\infty}\frac{\bar{x}-z}{4\pi\bar{J}^{3/2}} e^{-\frac{(\bar{x}-z)^2}{4\bar{J}}-(\frac{k^2}{4}+\gamma)\bar{J}-\frac{k}{2}(\bar{x}-z)}\int^{\tau}_{n\bar{J}}\frac{{{C_{A}(z,s)}}}{\sqrt{\tau-s}}e^{(\frac{k^2}{4}+\gamma)(\tau-s)}dsdz\\
-\frac{1}{\pi}\int^{\tau}_{n\bar{J}} \frac{e^{-(\frac{k^2}{4}+\gamma)(\tau-s)}}{\sqrt{\tau-s}} \int^{\sqrt{\bar{J}}}_{\sqrt{s-n\bar{J}}}e^{-(\frac{k^2}{4}+\gamma)t^2} \big[(\frac{k^2}{4}+\gamma)W_n(s-t^2) +W^{'}_n(s-t^2)\big]dtds.
\end{multline*}
There are several points that we should remark here. First, once $W(\tau)$ is found, the price of an American-style Parisian down-and-in call option can then be calculated straightforwardly by means of (\ref{1.3.7}) and (\ref{1.3.8}). The calculation procedure for an American-style Parisian down-and-in call option is very similar to that for a European Parisian up-and-out call as presented in \cite{zhu2013pricing}, except that we replace the value of the European vanilla option by the numerical value of its American counterpart, which is well documented in the literature. For simplicity, we do not present the calculation procedure in this paper.  Second, from the above solution for an American-style Parisian down-and-in call option, we can immediately derive a closed-form solution for European-style Parisian down-and-in call option by replacing the value of the American vanilla call option in the above formulae of $V_1, V_2, W$ by the value of an European call. Third, using American-style Parisian put-call symmetry  as in \cite{chesney2006american}, the solution for an Parisian knock-in put option can be derived from the call counterpart.

\section{Conclusion}
\vspace{-0.2cm}

In this paper, a simple analytical solution for American-style Parisian down-and-in call options is derived. This analytical solution can also be considered in a closed form if we suppose the value of embedded American vanilla call is known in advance. A key step of our approach is to apply the ``moving window" technique developed in \cite{zhu2013pricing} to simplify the pricing domain, and consequently reduce a 3-D problem to two coupled 2-D systems. Our solution procedure can be easily extended to other types of American-style Parisian knock-in options.

%In this paper, a closed-form analytical solution for American-style Parisian down-and-in call options is obtained for the first time. Realizing that there is no free boundary before the ``knock-in" feature has taken place is a key step that has helped us to find an elegant way to price American-style Parisian Down-and-in Call options. The solution technique developed in \cite{zhu2013pricing} is adopted to price American-style Parisian Down-and-in Call options and we obtain the resulting formula written in the form of four double integrals.

%\bibliographystyle{apalike}
%\linespread{0.1}
%{\small \bibliography{review}}

\end{document}